\documentclass[twocolumn,superscriptaddress]{revtex4-1}
\usepackage{color,soul,upgreek}

\usepackage{amsmath}    % need for subequations
\usepackage{graphicx}   % need for figures
\usepackage{verbatim}   % useful for program listings
\usepackage{color}      % use if color is used in text
\usepackage{subfigure}  % use for side-by-side figures
\usepackage{hyperref}   % use for hypertext links, including those to external documents and URLs
\usepackage{color}
\usepackage{amsmath}
\begin{document}
\title{Phase-Insensitive Scattering of Terahertz Radiation}
\author{Mihail \surname{Petev}}
\affiliation{School of Engineering and Physical Sciences, SUPA, Heriot-Watt University, Edinburgh EH14 
4AS, UK}
\affiliation{Max Planck Institute for the Science of Light (MPL), D-91058 Erlangen, Germany}
\author{Niclas \surname{Westerberg}}
\affiliation{School of Engineering and Physical Sciences, SUPA, Heriot-Watt University, Edinburgh EH14 
4AS, UK}
\author{Eleonora  \surname{Rubino}}
\affiliation{School of Engineering and Physical Sciences, SUPA, Heriot-Watt University, Edinburgh EH14 
4AS, UK}
\author{Daniel \surname{Moss}}
\affiliation{School of Engineering and Physical Sciences, SUPA, Heriot-Watt University, Edinburgh EH14 
4AS, UK}
\affiliation{Department of Physics and Solid State Institute, Technion, Haifa 32000, Israel}
\author{Arnaud \surname{Couairon}}
\affiliation{Centre de Physique Th\'eorique CNRS, \'Ecole Polytechnique, F-91128 Palaiseau, France}
\author{Fran\c{c}ois \surname{L\'egar\'e}}
\affiliation{INRS-EMT, 1650 Blvd. Lionel-Boulet, Varennes, QC J3X 1S2, Canada}
\author{Roberto \surname{Morandotti}}
\affiliation{INRS-EMT, 1650 Blvd. Lionel-Boulet, Varennes, QC J3X 1S2, Canada}
\affiliation{Institute of Fundamental and Frontier Sciences, University of Electronic Science and Technology of China, Chengdu 610054, China}
\affiliation{National Research University of Information Technologies, Mechanics and Optics, St. Petersburg 197101, Russia}
\author{Daniele \surname{Daniele}}
\email[E-mail: ]{D.Faccio@hw.ac.uk}
\affiliation{School of Engineering and Physical Sciences, SUPA, Heriot-Watt University, Edinburgh EH14 4AS, UK}
\author{Matteo \surname{Clerici}}
\email[E-mail: ]{matteo.clerici@glasgow.ac.uk}
\affiliation{School of Engineering, University of Glasgow, Glasgow G12 8LT, UK}

\begin{abstract}
The nonlinear interaction between Near-Infrared (NIR) and Terahertz pulses is principally investigated as a means for the detection of radiation in the hardly accessible THz spectral region. Most studies have
~targeted second-order nonlinear processes, given their higher efficiencies, and only a limited number have addressed third-order nonlinear interactions, mainly investigating four-wave mixing in air for broadband THz detection.
We have studied the nonlinear interaction between THz and NIR pulses in solid-state media (specifically diamond), and we show how the former can be frequency-shifted up to UV frequencies by the scattering from the nonlinear polarisation induced by the latter. Such UV emission differs from the well-known \emph{electric field-induced second harmonic} (EFISH) one, as it is generated via a \emph{phase-insensitive} scattering, rather than a sum- or difference-frequency four-wave-mixing process.
\end{abstract}
\maketitle

\section{Introduction}
 The far-infrared spectral region suffered for years from the lack of adequate sources and detectors---a fact that led the community to identify this evident underdevelopment as the \emph{THz Gap}. The potential impact of THz photonics on fields such as bio-imaging, pharmaceutical, chemical identification, and security stimulated the research on this topic~\cite{Federici2005a,Wallin2009}, and the \emph{THz Gap} in now almost closed~\cite{Tonouchi2007}. To this end, the development of practical THz sources and detectors witnessed in the last two decades played a crucial role. Many techniques for THz detection rely on the nonlinear red interaction of the THz radiation with visible or NIR probe pulses. For instance, the electro-optical effect in second-order nonlinear crystals is at the core of several time-domain spectroscopy setups~\cite{Zhang2010c}. Alternatively, third-order nonlinear interactions in centrosymmetric media such as \emph{electric field-induced second harmonic generation} (EFISH) have been proposed as means to detect infrared radiation~\cite{Ohlhoff1996, Nahata1998}.  

A breakthrough in broadband  time-domain THz spectroscopy was achieved utilising the low dispersion of air to perform THz coherent detection~\cite{Dai2006}. This technique, later called \emph{air-biased coherent detection} (ABCD)~\cite{Karpowicz2008}, relies on the EFISH process, which can be interpreted---in the framework of four-wave-mixing---as the beating between sum- and difference-frequency generation processes~\cite{Clerici2013g}. Such beating leads to a phase- and amplitude-modulated signal at the second harmonic of the optical probe. Overlapping this signal to a coherent field at the second-harmonic of the probe pulse, either induced by spectral broadening of the high intensity probe in air~\cite{Dai2006}, by electric field-induced second harmonic generation of the probe via an external DC electric field~\cite{Karpowicz2008}, or by employing a second-order nonlinear crystal~\cite{Li2015} leads to coherent detection of the THz field. In a recent publication, we investigated such a four-wave-mixing process in condensed matter (specifically, diamond), demonstrating counter-propagating phase-matched THz--NIR (near infrared) mixing~\cite{Clerici2013c}.

In this letter, we study the emission spectrum of the interaction between a THz single-cycle field and a NIR pump pulse mediated by the third-order nonlinearity. While at low pump intensities we observe the expected EFISH spectral signature, at high intensities, we observe the appearance of an additional feature in the UV spectral region that hints at the onset of a nonlinear process different from the EFISH. We performed a numerical study of the nonlinear waves dynamics, and we identified the unexpected signal as the result of a\emph{ phase-insensitive scattering} (PI) of the THz from the intense NIR pump~\cite{Efimov2005,Efimov2006}. These results are in keeping with our previous prediction for the scattering from an effective moving dispersive medium~\cite{Petev2013}.

\section{Experiments}
\begin{figure}[t!]
\centering
\includegraphics[width=\columnwidth]{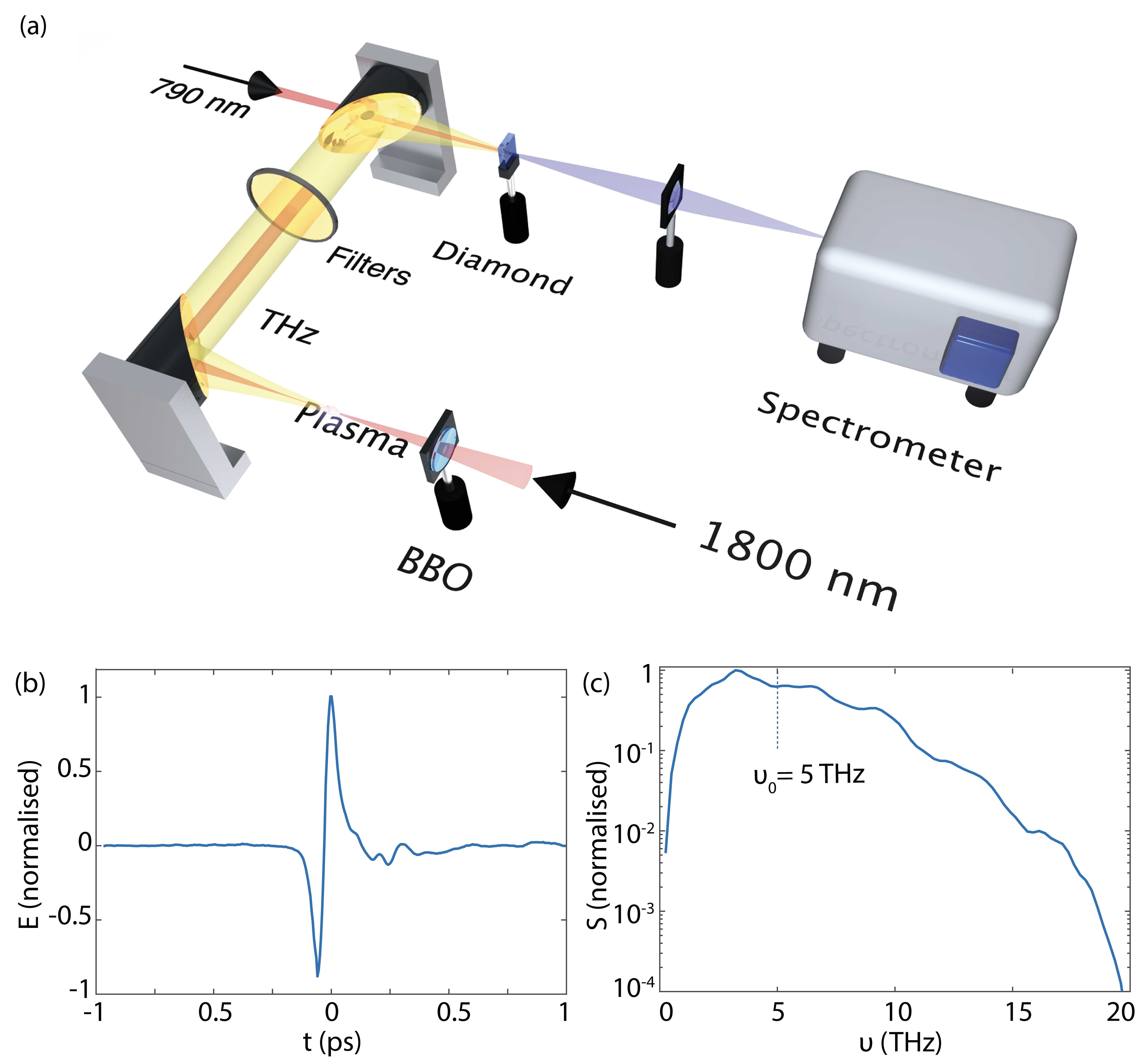}
\caption{{\color{black} (\textbf{a})} Experimental setup. A broadband (20 THz) seed pulse is generated by field ionisation of nitrogen in an asymmetric field composed of $1800$~nm and $900$~nm radiation. {\color{black} The THz electric field recorded via air-biased coherent detection (ABCD) and its spectrum are shown in (\textbf{b},\textbf{c}), respectively}.  The  THz field is collimated by a gold-coated off-axis parabolic mirror and filtered by two gold mesh long pass filters in order to remove every frequency component above $20$~THz. The THz pulse is then focused in a diamond single crystal sample collinearly, and is temporally overlapped with an intense~$790$~nm pump pulse. The generated UV radiation is collected by a lens and detected by an~imaging spectrometer coupled to a charge-coupled device (CCD).}
\label{setup}
\end{figure}
 We experimentally investigated the nonlinear interaction between intense NIR pulses and THz fields in diamond by recording the spectrum resulting from their nonlinear mixing. The setup employed is sketched in Figure~\ref{setup}a. The NIR pump pulses are delivered at a $100$~Hz repetition rate by a Ti:sapphire laser (\emph{Advanced Laser Light Source, ALLS }, Thales), with $790$~nm central wavelength and $40$~fs full-width at half-maximum duration. The seed pulse at THz frequencies is generated via two-colour gas ionisation driven by a mid-infrared pump at $1.8$ $\upmu$m and its second harmonic, as described in Ref.~\cite{Clerici2013d}.  A set of long pass filters with a cutoff at $20$~THz and $10^{4}$ isolation was employed to remove the high-frequency components from the THz pulse (QMC Instruments).
 
We chose diamond for the nonlinear mixing, mainly because it features high optical transmission over a large spectral region, covering all the  bands involved in our experiments ({i.e.}, NIR, UV,  and THz). The THz pulse is focused with a $2$~inch diameter, $2$~inch reflected focal length, gold-coated, $90$~deg off-axis parabolic mirror onto a 500~$\upmu$m thick, $\left<100\right>$-cut, single crystal diamond sample (Element-6). The THz beam is characterised by scanning the focused region with a THz camera (PV-320, Electrophysics). The mode profile is that of a broadband Bessel--Gauss beam with a $\sim$85~$\upmu$m width, Gaussian apodization, and $13$~deg cone-angle. The THz field temporal profile at the parabolic focus was measured with ABCD~\cite{Karpowicz2008}, and resulted in a single cycle sine wave pulse {\color{black} (see  Figure~\ref{setup}b}) with $\sim$90~fs duration and {\color{black} $\sim$5~THz carrier frequency ($\simeq$60~$\upmu$m wavelength), see Figure~\ref{setup}c}. The THz pulse energy was measured by a calibrated pyroelectric detector (Molectron, Coherent), and was $\sim$1~$\upmu$J. The~NIR pump pulse is tightly focused with a $f=125$~mm fused-silica lens, and is overlapped in space and time to the THz field inside the diamond sample.

\begin{figure}[t!]
\centering
\includegraphics[width=\columnwidth]{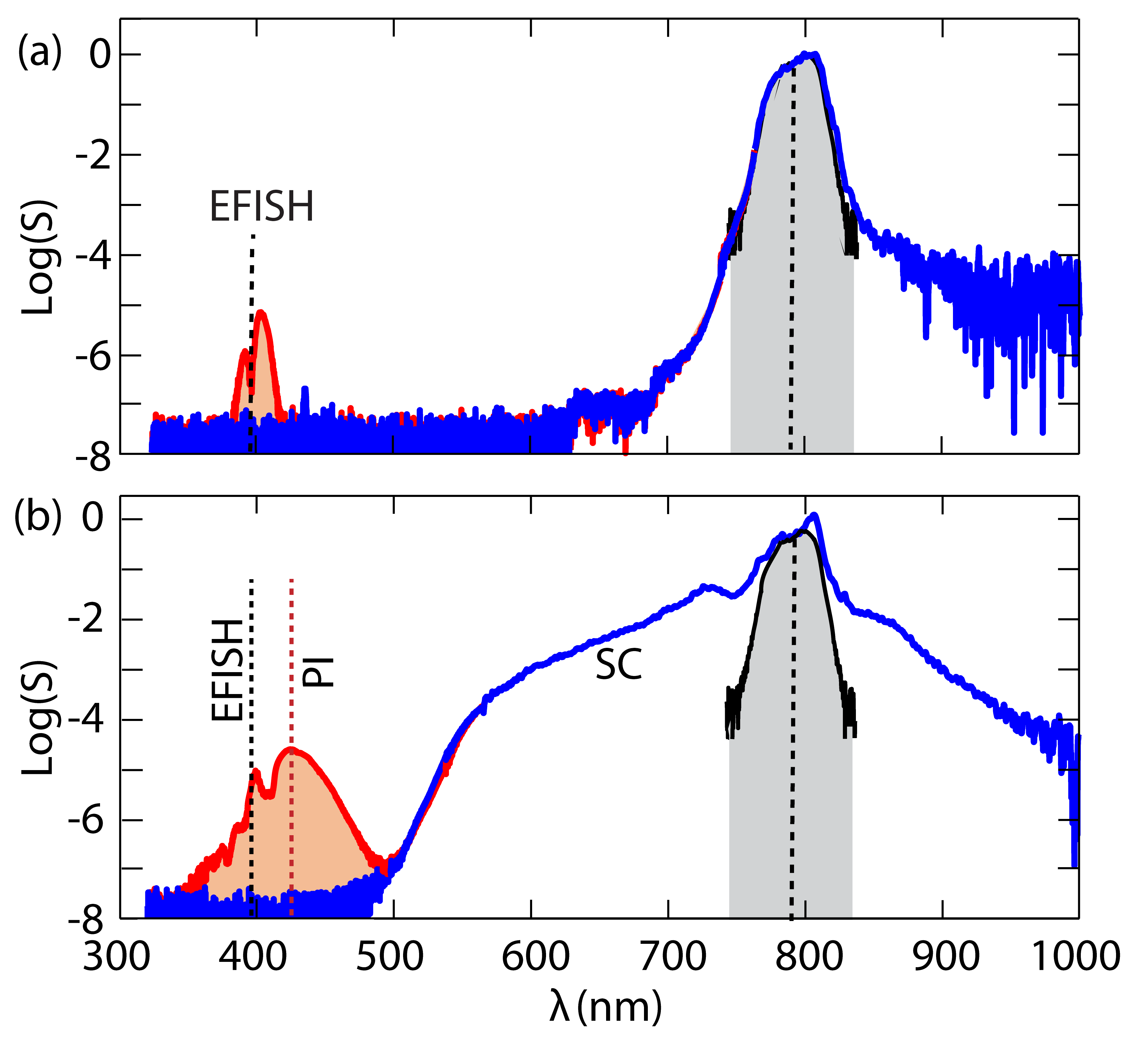}
\caption{(\textbf{a}) Experimental measurement showing the electric field-induced second harmonic (EFISH) generation signal typically observed for low pump intensities (red, shaded curve, {\color{black} S stands for power spectrum)}. The blue curve shows the pump pulse spectrum after the propagation in the crystal without any injected THz seed. The black shaded curve shows a section of the input spectrum (a very limited spectral reshaping of the pump pulse is evident); (\textbf{b}) as in (a), but for a higher pump pulse intensity. The occurrence of an additional signal at $\sim$$430$ nm is clearly visible in the red shaded~curve.}
\label{exp_fig}
\end{figure}

The spectrum of the radiation at the output of the crystal was measured with an imaging spectrometer (Newport MS260i) coupled to a charge-coupled device (CCD) camera (QSI 620). Different neutral density filters have been employed while acquiring data on different portions of the spectrum to enhance the overall dynamic range. The measured spectra recorded for low $3$~$\upmu$J and high $10$~$\upmu$J pump energies are shown in  Figure~\ref{exp_fig}a,b, respectively. For the low pump energy case, the spectrum is what  would be
~expected from an EFISH process, as can be appreciated considering the spectrally-resolved measurements of the nonlinear wave-mixing in air between the same THz and NIR pulses reported in Ref.~\cite{Clerici2013g}. The red curve in Figure~\ref{exp_fig}a shows two spectral lobes around the second harmonic wavelength of the NIR pump, according to:
\begin{equation}
 \omega_{\textrm{EFISH}}^{\textrm{SF/DF}}=2\omega_p\pm\omega_{\textrm{seed}},
 \label{EFISH}
 \end{equation}
where SF and DF stand for sum- and difference-frequency, respectively, $\omega_p=2\,\pi\,c/\lambda_p$, $\lambda_p=790$~nm, and $\omega_{\textrm{seed}}=2\,\pi\,$5~THz. {\color{black} We note that the two measured peaks are $\simeq$10~nm apart, whereas the spectral shift predicted for a $5$~THz carrier frequency pulse is $\simeq$5 nm. This difference is likely to be due to the limited THz detection bandwidth accessible with the long $40$~fs probe pulse, and to phase matching considerations that are not included in Equation~\eqref{EFISH}}.
Notably, at these low input energies and in the absence of an input THz seed, the pump pulse undergoes minimal nonlinear reshaping, and the output spectrum (blue curve) is still close to that of the input spectrum (black curve in Figure~\ref{exp_fig}a).

On the other hand, when the pump intensity is increased, a strong spectral reshaping occurs, as can be seen by the onset of supercontinuum generation, indicated by SC on the blue curve in Figure~\ref{exp_fig}b. Supercontinuum generation is a signature of strong nonlinear effects, and the blue-shift indicates significant self-steepening~\cite{Couairon2007a}. In this regime, a second, more intense peak centred at $\sim$$430$~nm appears, and in the following, we provide our interpretation of its physical origin. {\color{black} We~note that in the high-energy case, the SF and DF peaks are not spectrally separated, as a consequence of the strong spectral broadening of the intense pump field.}
In Ref.~\cite{Petev2013}, we have anticipated that the scattering of a THz field from the refractive index induced by the shock-front on an intense NIR pulse propagating in diamond would result in an emission in the UV part of the spectrum. In the following section, we show that the recorded $\sim$$430$~nm peak fits with the prediction of that model. Before that, we exclude some of the more trivial effects, such as EFISH and Raman scattering. 

\subsection{{\color{black} Excluded Possible Origins of the UV Signal}}

\begin{itemize}
\item{{\bf {\color{black} Fluorescence.}}} {\color{black} We note that although diamond exhibits fluorescence in the analysed UV spectral regions, all the recorded UV signals only appear when the THz field is spatially and temporally overlapped to the $790$~nm intense pulse. This observation rules out the possibility that the $\simeq$$430$~nm signal is originating from a fluorescence process in diamond.}
\item{{\bf {\color{black} Residual THz pump field.}}} {\color{black} The multimesh filters used to remove the $1800$~nm field from the THz beam path provide more than $10^4$ extinction. Given the low $<$10$^{-4}$ efficiency of the THz generation process, it is essential to exclude the possibility that a residual $1800$~nm field contributes to the onset of the $\simeq$$430$~nm signal. To this end, we have first inserted a $1$~mm thick $\textrm{CaF}_2$ window in the THz beam path and searched for evidence of UV radiation. In this case, no EFISH or $430$~nm signal were observed. Since $\textrm{CaF}_2$ is transparent to IR radiation but opaque to THz, we concluded that THz is essential to generate the $430$~nm signal. Further, we observed that the UV radiation was visible upon removing the $\textrm{CaF}_2$ window and inserting a paper filter, which both absorbs and scatters the $1800$~nm radiation. We therefore exclude any possible role of a residual $1800$~nm field in the generation of the $430$~nm signal.}
\item{{\bf EFISH.}} We note that the $430$~nm peak cannot be explained by the EFISH mechanism described above. Indeed, its large shift from $395$~nm---the wavelength of the pump second harmonic---cannot be justified by the broadening of the NIR pump pulse spectrum. 
\item{{\bf Raman.}} Diamond is a well-known Raman-active crystal and has a large $\Delta_R\sim1350$~cm$^{-1}$ shift. The Raman-mediated wave mixing, $\omega_p+\omega_p-2\,\pi\,c\,\Delta_R$, gives a product at $417$ nm \mbox{($c$ is the speed of light)}. This process is independent of the presence of the THz field, whereas in our measurement, the $430$~nm signal only appeared when the THz field was injected in the crystal. Another Raman-mediated four-wave-mixing process that may lead to a signal around $417$~nm is $\omega_p+(\omega_p-2\,\pi\,c\,\Delta_R)\pm \omega_{\textrm{seed}}$. This would require the presence of a Raman peak at  $\omega_p-2\,\pi\,c\,\Delta_R$, corresponding to a $\sim$$884$~nm wavelength, which does not appear in the recorded spectra. {\color{black} Finally, we note that the EFISH signal has energies $<$$1$~pJ, and is too low to directly excite a Raman red-shifted peak.}
\end{itemize}

\section{Model}

  Here we consider the experimental condition where a weak seed and a strong pump co-propagate in a dispersive medium. The temporal dynamics of the two-pulse interaction can be investigated within the framework of the First Born approximation (e.g., Ref~\cite{Kolesik2005}). In this case, the result of the interaction is evaluated considering the scattering of the seed pulse from a potential, which is determined by the nonlinear refractive index modification induced by the pump pulse via the Kerr nonlinearity. This description can also be applied in a guided wave geometry~\cite{Kolesik2010}, and if the two pulses have different frequencies, the weak one may catch up and scatter off the refractive index induced by the intense one.
Interestingly, if the refractive index perturbation is such that the seed pulse cannot overcome it, the interaction mimics the physics at an event horizon. This analogy holds for time-invariant pump pulses, such as solitons in an optical fibre~\cite{Philbin2008a,Webb2014}. However, the horizon physics can be observed in a wider context, such as in optical filamentation~\cite{Belgiorno2010}.

The discussed scattering process leads to a shift of the probe frequency, and is a relevant mechanism for the spectral broadening and the supercontinuum generation at the core of the nonlinear guided-optics research~\cite{Agrawal2001,Dudley2006a}. Within this framework, frequency conversion consequent to the interaction of a weak field with a strong solitonic pump has previously been investigated, and is termed \emph{phase-insensitive scattering}~\cite{Gordon1992,Yulin2004,Efimov2004,Skryabin2005,Efimov2005,Efimov2006}.

The scattering process is favoured for those spectral components that keep the same phase of the input seed across the interaction. This is determined by the momentum conservation:
\begin{equation}
 k(\omega_\textrm{scatter})-\omega_\textrm{scatter}/v_p=k({\omega_\textrm{seed}})-\omega_\textrm{seed}/v_p,
 \label{XP-RR}
\end{equation}
where $k$ is the wavevector, $\omega_\textrm{seed}$ is the weak seed frequency, $S$ indicates the scattered quantities, and the \emph{scatterer} velocity is that of the pump pulse, $v_p$. Considering the dispersion of diamond $k(\omega)$ from Ref.~\cite{Zaitsev2001a} and the linear input conditions of the experiment,  $v_p=\left[d\,k(\omega)/d\,\omega|_{\omega=\omega_p}\right]^{-1}={\color{black}0.1228}$~$\upmu$m/fs,  Equation \eqref{XP-RR} is satisfied for $\lambda_\textrm{S}=2\,\pi\,c/\omega_\textrm{scatter}\sim{\color{black}460}$~nm. This result is obtained neglecting any nonlinear effect that the pump pulse undergoes during the propagation in the diamond, yet the predicted wavelength for the maximum scattering efficiency is remarkably close to the  anomalous $\sim430$~nm signal measured in the experiment above. To refine this prediction, we performed numerical simulations with a code that solves the nonlinear Maxwell's equations in \mbox{(1D + 1)} for the total field ${\cal{E}}={\cal{E}}_{p}+{\cal{E}}_{\textrm{seed}}$, using the Pseudospectral Space Domain (PSSD) algorithm~\cite{Tyrrell2005}. Note that ${\cal{E}}$ refers to the complex electric field.

\subsection {Nonlinear Maxwell equation Via Pseudospectral Space Domain Algorithm (1D + 1) }
  \begin{figure}[t!]
\centering
\includegraphics[width=\columnwidth]{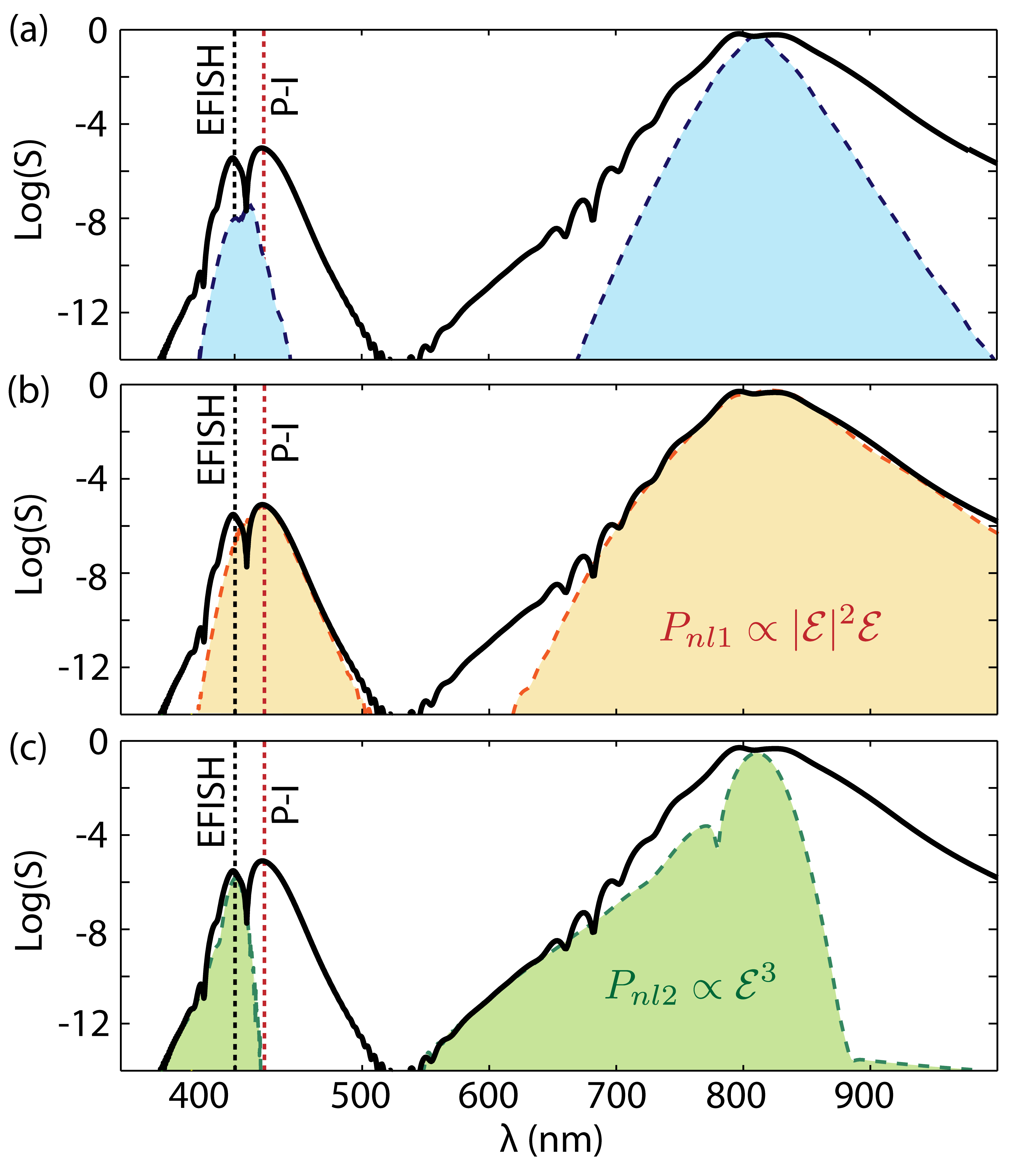}
\caption{(\textbf{a}) Numerically simulated spectra for the total (pump and THz) field---spectra are shown at the sample output ($z=5$ mm). The light blue shaded area is for low pump intensity, while the thick black curve is for high pump intensities {\color{black} (S stands for power spectrum)}. (\textbf{b}) Same as in (a), but comparing---for the high energy case---the results from a full simulation (black solid curve) with those from a simulation accounting only for the $P_{nl1}\propto\textrm{Re}\left[ |{\cal{E}}|^2{\cal{E}}\right]$ (red shaded curve). (\textbf{c}) Same as (b), but comparing the full simulation (black solid curve) with one including only the $P_{nl2}\propto\textrm{Re}\left[ {\cal{E}}^3\right]$ term (green shaded curve). PI: phase insensitive scattering.}
\label{fig3}
\end{figure}
  
\noindent The PSSD approach solves the Maxwell equations for the electric and magnetic fields, considering the fields as functions of time, performing the spatial derivatives in direct space and the temporal derivatives in the Fourier domain. We considered a nonlinear polarisation $P_{nl}$ written in a fashion identical to that presented in Ref.~\cite{Conforti2013} that
~explicitly separates  $P_{nl}$ into two terms: \scalebox{.95}[1.0]{$P_{nl}=P_{nl1} + P_{nl2}$}. $P_{nl1}\propto\textrm{Re}\left[ |{\cal{E}}|^2{\cal{E}}\right]$ is responsible for the \emph{phase-insensitive} scattering process as well as the difference frequency generation components of the EFISH. $P_{nl2}\propto\textrm{Re}\left[ {\cal{E}}^3\right]$ accounts for third harmonic generation and other four-wave mixing effects, including the sum-frequency generation term of the EFISH. Combined together, these terms describe the full plethora of third-order nonlinear effects. 

Figure~\ref{fig3}a shows the result of these simulations. The light blue shaded area shows the output spectrum for a low pump energy. As can be seen, a weak EFISH signal is observed around the pump second harmonic at $\sim$$395$~nm, in keeping with our experimental observations shown in Figure~\ref{exp_fig}a. Here, two photons of the intense pump pulse perform a sum and a difference frequency generation process together with a THz photon, hence generating fields with frequencies centred at the pump second harmonic. The momentum conservation relation for the EFISH process is $k_\textrm{EFISH}=2k_p + k_{\textrm{seed}}$, while the energy conservation relation is in Equation \eqref{EFISH}.
{\color{black} We stress that these simulations are performed in $1D+1$, and therefore are not expected to capture all the details of the measured spectra. Nonetheless, at} high pump input energies, {\color{black} they predict the appearance of a broad supercontinuum (SC) which indicates the onset of a shock, and a second peak at  $\sim$$430$~nm, in addition to the EFISH peak at $\sim$$395$~nm}, as shown by the black curve in Figure~\ref{fig3}a. {\color{black} The spectral position of these peaks agrees} with the experiments shown in Figure~\ref{exp_fig}b.
  To gain insight into the physical mechanism responsible for the $430$~nm emission, we performed two sets of simulations for the case of high-energy  pump. The first (shown in Figure~\ref{fig3}b) is performed only including the term $P_{nl1} $. This term is expected to lead to the generation of the red-shifted part of the EFISH, as well as of the scattered radiation, and indeed matches well the long-wavelength lobe of the UV signal. A second simulation is performed considering only $P_{nl2}$ (as shown in Figure~\ref{fig3}b), and as expected, matches the blue-end part of the UV spectrum. From these simulations, we can infer that the red-shifted peak observed both in the simulation and in the experiments at $\sim$$430$~nm is due either to the difference frequency generation terms of four-wave mixing, or to the \emph{phase-insensitive} scattering process. To determine which of the two terms is indeed contributing to the measured spectrum, we performed another set of simulations, this time dropping the four-wave mixing terms. In addition, we consider a more general 2D + 1 scenario where only cylindrical symmetry around the propagation axis is assumed.
\subsection{Coupled Nonlinear Envelope Equations}
\begin{figure}[t!]
\centering
\includegraphics[width=\columnwidth]{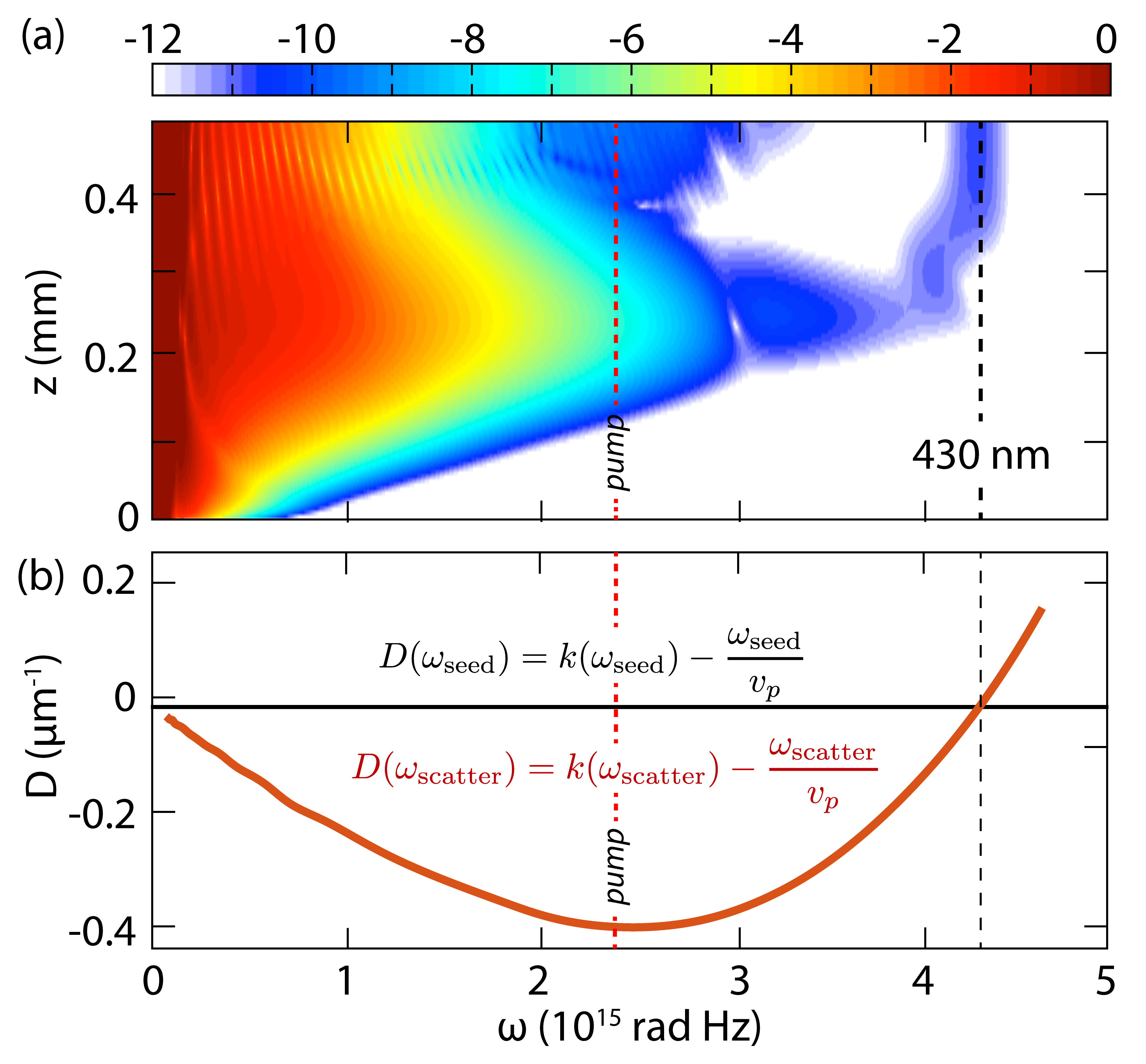}
\caption{{\color{black} (\textbf{a}) Numerically simulated evolution of the seed field spectrum along the full $5$~mm propagation distance based on Equation~\eqref{NEEh}  (colormap is in logarithmic scale).  (\textbf{b})  Dispersion, $D(\omega_{\textrm{scatter}})=k(\omega_{\textrm{scatter}})-\omega_{\textrm{scatter}}/v_p$, for diamond calculated for the speed of the shock front, $v_p=\sim1.223 \times 10^8$ m/s (red curve). The horizontal line shows $D(\omega_\textrm{seed})=k(\omega_\textrm{seed})-\omega_\textrm{seed}/v_p$: the intersections with $D(\omega_{\textrm{scatter}})$ gives the spectral location of the phase-insensitive scattering peak. The blue-shaded box shows the THz input spectrum at $1/e^2$. For the sake of clarity, the spectral position of the $790$~nm pump pulse is also shown with a vertical red dashed line.}}
\label{fig4}
\end{figure}
  We use a numerical model that relies on two coupled nonlinear envelope equations (NEE) describing the propagation of the intense pump laser pulse with frequency $\omega_p$ coupled with a second THz pulse at frequency $\omega_{\textrm{seed}}$.  The coupled NEE are written for the Fourier-transformed envelopes of the fundamental pulse $\hat{\cal E}_{p}(\delta \omega_p, r, z)$ and the THz pulse $\hat{\cal E}_{{\textrm{seed}}}(\delta \omega_{\textrm{seed}}, r, z)$, and read as \cite{Couairon2011a}:
\begin{align}
\frac{\partial \hat{\cal E}_{p}}{\partial z} =& 
{i} \left[\frac{\nabla^2_{\perp}}{2k_p(\omega)} + {\cal K}_p\right]\hat{\cal E}_{p} +
 {i} \frac{n_0 n_2\omega^2}{k_p(\omega) c^2}\widehat{|{\cal E}_{p}|^2{\cal E}_{p}}, \label{NEE0} \\
\frac{\partial \hat{\cal E}_{\textrm{seed}}}{\partial z} =& 
{i} \left[\frac{\nabla^2_{\perp}}{2k_{\textrm{seed}}(\omega)} + {\cal K}_{\textrm{seed}}(\omega) - \Delta{k}\right]\hat{\cal E}_{\textrm{seed}}+\nonumber  \\
&+{i} \frac{2n_0 n_2\omega^2 }{k_{\textrm{seed}}(\omega) c^2}  \widehat{|{\cal E}_{p}|^2{\cal E}_{\textrm{seed}}} \label{NEEh}
\end{align}
where $n_2=1.2 \times 10^{-15}$ cm$^2$/W  \cite{Boyd2008b} , $k_p(\omega)= k(\omega_p) + k_p' \delta \omega_p$,  ${\cal K}_p(\omega)= k(\omega)-k_p(\omega)$, and similarly for the {\color{black} seed,} THz field, $k_{\textrm{seed}}(\omega) = k(\omega_{\textrm{seed}}) + {k^{\prime}_{\textrm{seed}} \delta{\omega_{\textrm{seed}}}}$, ${\cal K}_{\textrm{seed}}(\omega)= k(\omega)-k_{\textrm{seed}}(\omega)$, $\Delta k = k_0' \delta \omega_{\textrm{seed}} - k(\omega_{\textrm{seed}})$. The \emph{hat} ($\hat{\cdot}$) symbol denotes the time-domain Fourier transform. The frequency variable $\delta{\omega}$  is fixed for both pulses by the numerical grid, and represents the departure from $\omega_p$ and $\omega_{\textrm{seed}}$ in Equations~\eqref{NEE0} and~\eqref{NEEh}, respectively. We note that this model does not account for effects such as the four-wave mixing between the seed THz pulse and the NIR pump, and hence does not capture the EFISH process. It is therefore well-suited to evidence the \emph{phase-insensitive} scattering~effects.

As input for our simulation, we use a $790$~nm pump pulse with duration of $40$~fs and input energy of 10~$\upmu$J. The probe has a wavelength of $60$~$\upmu$m, duration of $90$~fs, and energy of $2$~$\upmu$J, in~accordance with the experimental values. 

We introduced a relative delay of $110$~fs for the THz field in order to compensate for the difference in group velocities of the two pulses. This delay was optimised so that the THz pulse catches up with the pump pulse after a propagation distance of $\sim300$~$\upmu$m, corresponding to a position where the pump pulse generated the steepest shock front, which increases the efficiency of the resonant scattering process \cite{Rubino2012,Conforti2014}.

Figure~\ref{fig4}a shows in detail the spectral evolution of the THz pulse over the full propagation distance. At $z=0.3$~mm,  the interaction with the pump pulse induces \emph{phase-insensitive} scattering. The scattering peak in Figure~\ref{fig4}a occurs at a frequency corresponding to $430$~nm, in agreement with the simulations performed with the PSSD algorithm, and most importantly, with the experimental observations. This supports the interpretation of the observed peak at $430$~nm as a result of \emph{phase~insensitive} scattering of the THz seed off the refractive index perturbation induced by the intense pump.
 {\color{black} It is worth noting that the numerical simulation in Figure~\ref{fig4}a does not include cross-phase modulation induced by the intense $790$~nm pulse on the $\simeq$$395$~nm EFISH signal among the captured physical effects. Furthermore, the amplitude of the~$\simeq$$430$~nm signal is comparable to that of the EFISH, and we are therefore confident to exclude such cross-phase modulation process from the possible sources of the $\simeq$$430$~nm emission.}

To further support this interpretation, we evaluated the velocity of the shock front on the pump pulse as resulted from the 2D + 1 NEE simulations. At the propagation distance at which the scattering occurs (i.e., at $\sim$$300$~$\upmu$m), the front speed is $\sim$$0.1223$~$\upmu$m/fs. Considering this as the value for the speed of the scattering object, the position of the phase-insensitive scattering peak can be predicted from Equation~\eqref{XP-RR}, and is graphically shown in Figure~\ref{fig4}a, where we plot the dispersion \mbox{$D=k(\omega_{\textrm{scatter}})-\frac{\omega_{\textrm{scatter}}}{v_p}$} (red curve), while the black horizontal line is by $D=k(\omega_\textrm{seed})-\omega_\textrm{seed}/v_p$. There is an excellent correspondence between the predicted and numerically observed spectral peaks, thus confirming our interpretation of the $430$~nm peak as \emph{phase-insensitive} scattering.

\section{Conclusions}
 Frequency conversion of THz radiation in the UV spectral range can be accomplished via EFISH, and it provides a mean for the detection of far-infrared radiation. Here we have shown that, under appropriate conditions, an additional effect contributes to the frequency conversion; namely, phase-insensitive scattering. This effect belongs to the cross-phase modulation family rather than to the frequency mixing processes that accompany the third-order nonlinearity. 

\vspace{6pt}
%%%%%%%%%%%%%%%%%%%%%%%%%%%%%%%%%%%%%%%%%%
\acknowledgments{D.F. acknowledges financial support from the European Research Council under the European Union's Seventh Framework Programme (FP/2007-2013)/ERC GA 306559 and EPSRC (UK, Grant EP/J00443X/1). M.C. acknowledges the support from the People Programme (Marie Curie Actions) of the European Union's Seventh Framework Programme (FP7/2007-2013) GA 299522 and from EPSRC (EP/P009697/1). N.W. acknowledges support from the EPSRC CM-CDT Grant No. EP/L015110/1. R.M. gratefully acknowledges support by NSERC (Strategic Grant Program), MESI (PSR-SIIRI Grant Program) and FQRNT (Equipe Grant Program) in Canada. The authors wish to acknowledge the technical support from the staff of the ALLS (Advanced Laser Light Source) facility, from  M. Peccianti and from B. Schmidt. 
M.C. gratefully acknowledges the support of B.L.M. SpA. The dataset which underpins this publication is available at \url{http://dx.doi.org/10.5525/gla.researchdata.373}.\\

All the authors equally contributed to the development of the work and the writing of the manuscript.\\

The authors declare no conflict of interest. The founding sponsors had no role in the design of the study; in the collection, analyses, or interpretation of data; in the writing of the manuscript, and in the decision to publish the results}

%%%%%%%%%%%%%%%%%%%%%%%%%%%%%%%%%%%%%%%%%%
%\authorcontributions{All the authors equally contributed to the development of the work and the writing of the manuscript.}

%%%%%%%%%%%%%%%%%%%%%%%%%%%%%%%%%%%%%%%%%%
%\conflictofinterests{The authors declare no conflict of interest. The founding sponsors had no role in the design of the study; in the collection, analyses, or interpretation of data; in the writing of the manuscript, and in the decision to publish the results.} 

%%%%%%
%\bibliographystyle{mdpi}%mdpi
%\bibliography{THzRR.bib}

\begin{thebibliography}{5}
%\providecommand{\natexlab}[1]{#1}

\bibitem[Federici \em{et~al.}(2005)Federici, Schulkin, Huang, Gary, Barat,
  Oliveira, and Zimdars]{Federici2005a}
Federici, J.F.; Schulkin, B.; Huang, F.; Gary, D.; Barat, R.; Oliveira, F.;
  Zimdars, D., THz imaging and sensing for security applications--Explosives,
  weapons and drugs. {\em Semicond. Sci. Technol.} {\bf 2005}, {\em
  20},~S266--S280.

\bibitem[Wallin \em{et~al.}(2009)Wallin, Pettersson, {\"{O}}stmark, and
  Hobro]{Wallin2009}
Wallin, S.; Pettersson, A.; {\"{O}}stmark, H.; Hobro, A. Laser-based standoff detection of explosives: A critical review. {\em Anal. Bioanal. Chem.} {\bf 2009}, {\em
  395},~259--274.

\bibitem[Tonouchi(2007)]{Tonouchi2007}
Tonouchi, M.
\newblock {Cutting-edge terahertz technology}.
\newblock {\em Nat. Photonics} {\bf 2007}, {\em 1},~97--105.

\bibitem[Zhang and Xu(2010)]{Zhang2010c}
Zhang, X.C.; Xu, J.
\newblock {\em {Introduction to THz Wave Photonics}}; Springer US: Boston, MA, USA,
  2010; pp. 1--246.

\bibitem[Ohlhoff \em{et~al.}(1996)Ohlhoff, Meyer, Lupke, Loffler, Pfeifer,
  Roskos, and Kurz]{Ohlhoff1996}
Ohlhoff, C.; Meyer, C.; Lupke, G.; Loffler, T.; Pfeifer, T.; Roskos, H.G.;
  Kurz, H.
\newblock {Optical second-harmonic probe for silicon millimeter-wave circuits}.
\newblock {\em Appl. Phys. Lett.} {\bf 1996}, {\em 68},~1699.

\bibitem[Nahata and Heinz(1998)]{Nahata1998}
Nahata, A.; Heinz, T.F.
\newblock {Detection of freely propagating terahertz radiation by use of
  optical second-harmonic generation}.
\newblock {\em Opt. Lett.} {\bf 1998}, {\em 23},~67.

\bibitem[Dai \em{et~al.}(2006)Dai, Xie, and Zhang]{Dai2006}
Dai, J.; Xie, X.; Zhang, X.C.
\newblock {Detection of Broadband Terahertz Waves with a Laser-Induced Plasma
  in Gases}.
\newblock {\em Phys. Rev. Lett.} {\bf 2006}, {\em 97},~103903.

\bibitem[Karpowicz \em{et~al.}(2008)Karpowicz, Dai, Lu, Chen, Yamaguchi, Zhao,
  Zhang, Zhang, Zhang, Price-Gallagher, Fletcher, Mamer, Lesimple, and
  Johnson]{Karpowicz2008}
Karpowicz, N.; Dai, J.; Lu, X.; Chen, Y.; Yamaguchi, M.; Zhao, H.; Zhang, X.C.;
  Zhang, L.; Zhang, C.; Price-Gallagher, M.; et al.
%  Fletcher, C.; Mamer, O.; Lesimple,  A.; Johnson, K.
\newblock {Coherent heterodyne time-domain spectrometry covering the entire
  “terahertz gap”}.
\newblock {\em Appl. Phys. Lett.} {\bf 2008}, {\em 92},~011131.

\bibitem[Clerici \em{et~al.}(2013)Clerici, Faccio, Caspani, Peccianti, Yaakobi,
  Schmidt, Shalaby, Vidal, L{\'{e}}gar{\'{e}}, Ozaki, and
  Morandotti]{Clerici2013g}
Clerici, M.; Faccio, D.; Caspani, L.; Peccianti, M.; Yaakobi, O.; Schmidt,
  B.E.; Shalaby, M.; Vidal, F.; L{\'{e}}gar{\'{e}},~F.; Ozaki, T.; et al.
%  Morandotti,  R.
\newblock {Spectrally resolved wave-mixing between near- and far-infrared
  pulses in gas}.
\newblock {\em New J. Phys.} {\bf 2013}, {\em 15},~125011.

\bibitem[Li \em{et~al.}(2015)Li, Seletskiy, Yang, and Sheik-Bahae]{Li2015}
Li, C.Y.; Seletskiy, D.V.; Yang, Z.; Sheik-Bahae, M.
\newblock {Broadband field-resolved terahertz detection via laser induced air
  plasma with controlled optical bias}.
\newblock {\em Opt. Express} {\bf 2015}, {\em 23},~11436--11443.

\bibitem[Clerici \em{et~al.}(2013)Clerici, Caspani, Rubino, Peccianti,
  Cassataro, Busacca, Ozaki, Faccio, and Morandotti]{Clerici2013c}
Clerici, M.; Caspani, L.; Rubino, E.; Peccianti, M.; Cassataro, M.; Busacca,
  A.; Ozaki, T.; Faccio, D.; Morandotti, R.
\newblock {Counterpropagating frequency mixing with terahertz waves in
  diamond}.
\newblock {\em Opt. Lett.} {\bf 2013}, {\em 38},~178--180.

\bibitem[Efimov \em{et~al.}(2005)Efimov, Yulin, Skryabin, Knight, Joly,
  Omenetto, Taylor, and Russell]{Efimov2005}
Efimov, A.; Yulin, A.V.; Skryabin, D.V.; Knight, J.C.; Joly, N.; Omenetto,
  F.G.; Taylor, A.J.; Russell, P.
\newblock {Interaction of an Optical Soliton with a Dispersive Wave}.
\newblock {\em Phys. Rev. Lett.} {\bf 2005}, {\em 95},~213902.

\bibitem[Efimov \em{et~al.}(2006)Efimov, Taylor, Yulin, Skryabin, and
  Knight]{Efimov2006}
Efimov, A.; Taylor, A.J.; Yulin, A.V.; Skryabin, D.V.; Knight, J.C.
\newblock {Phase-sensitive scattering of a continuous wave on a soliton}.
\newblock {\em Opt. Lett.} {\bf 2006}, {\em 31},~1624--1626.

\bibitem[Petev \em{et~al.}(2013)Petev, Westerberg, Moss, Rubino, Rimoldi,
  Cacciatori, Belgiorno, and Faccio]{Petev2013}
Petev, M.; Westerberg, N.; Moss, D.; Rubino, E.; Rimoldi, C.; Cacciatori, S.L.;
  Belgiorno, F.; Faccio, D.
\newblock {Blackbody Emission from Light Interacting with an Effective Moving
  Dispersive Medium}.
\newblock {\em Phys. Rev. Lett.} {\bf 2013}, {\em 111},~043902.

\bibitem[Clerici \em{et~al.}(2013)Clerici, Peccianti, Schmidt, Caspani,
  Shalaby, Gigu{\`{e}}re, Lotti, Couairon, L{\'{e}}gar{\'{e}}, Ozaki, Faccio,
  and Morandotti]{Clerici2013d}
Clerici, M.; Peccianti, M.; Schmidt, B.E.; Caspani, L.; Shalaby, M.;
  Gigu{\`{e}}re, M.; Lotti, A.; Couairon, A.; L{\'{e}}gar{\'{e}}, F.; Ozaki,
  T.; et al.
\newblock {Wavelength Scaling of Terahertz Generation by Gas Ionization}.
\newblock {\em Phys. Rev. Lett.} {\bf 2013}, {\em 110},~253901.

\bibitem[Couairon and Mysyrowicz(2007)]{Couairon2007a}
{Couairon, A.; Mysyrowicz, A. Femtosecond filamentation in transparent media. {\em Phys. Rep.} {\bf 2007}, {\em 441}, 47--189.}

\bibitem[Kolesik \em{et~al.}(2005)Kolesik, Wright, and Moloney]{Kolesik2005}
Kolesik, M.; Wright, E.M.; Moloney, J.V.
\newblock {Interpretation of the spectrally resolved far field of femtosecond
  pulses propagating in bulk nonlinear dispersive media}.
\newblock {\em Opt. Express} {\bf 2005}, {\em 13},~10729--10741.

\bibitem[Kolesik \em{et~al.}(2010)Kolesik, Tartara, and Moloney]{Kolesik2010}
Kolesik, M.; Tartara, L.; Moloney, J.V.
\newblock {Effective three-wave-mixing picture and first Born approximation for
  femtosecond supercontinua from microstructured fibers}.
\newblock {\em Phys. Rev. A} {\bf 2010}, {\em 82},~045802.

\bibitem[Philbin \em{et~al.}(2008)Philbin, Kuklewicz, Robertson, Hill, Konig,
  and Leonhardt]{Philbin2008a}
Philbin, T.G.; Kuklewicz, C.; Robertson, S.; Hill, S.; Konig, F.; Leonhardt, U.
\newblock {Fiber-Optical Analog of the Event Horizon}.
\newblock {\em Science} {\bf 2008}, {\em 319},~1367--1370.

\bibitem[Webb \em{et~al.}(2014)Webb, Erkintalo, Xu, Broderick, Dudley, Genty,
  and Murdoch]{Webb2014}
Webb, K.E.; Erkintalo, M.; Xu, Y.; Broderick, N.G.R.; Dudley, J.M.; Genty, G.;
  Murdoch, S.G.
\newblock {Nonlinear optics of fibre event horizons}.
\newblock {\em Nat. Commun.} {\bf 2014}, {\em 5},~4969.

\bibitem[Belgiorno \em{et~al.}(2010)Belgiorno, Cacciatori, Clerici, Gorini,
  Ortenzi, Rizzi, Rubino, Sala, and Faccio]{Belgiorno2010}
Belgiorno, F.; Cacciatori, S.L.; Clerici, M.; Gorini, V.; Ortenzi, G.; Rizzi,
  L.; Rubino, E.; Sala, V.G.; Faccio, D.
\newblock {Hawking Radiation from Ultrashort Laser Pulse Filaments}.
\newblock {\em Phys. Rev. Lett.} {\bf 2010}, {\em 105},~203901.

\bibitem[Agrawal(2001)]{Agrawal2001}
Agrawal, G.
\newblock {\em Nonlinear Fiber Optics};
\newblock {Academic Press: New York, NY, USA, 2001; p. 467.}

\bibitem[Dudley \em{et~al.}(2006)Dudley, Genty, and Coen]{Dudley2006a}
Dudley, J.M.; Genty, G.; Coen, S.
\newblock {Supercontinuum generation in photonic crystal fiber}.
\newblock {\em Rev. Mod. Phys.} {\bf 2006}, {\em 78},~1135--1184.

\bibitem[Gordon(1992)]{Gordon1992}
Gordon, J.P.
\newblock {Dispersive perturbations of solitons of the nonlinear
  Schr\"odinger equation}.
\newblock {\em J. Opt. Soc. Am.  B} {\bf 1992}, {\em
  9},~91--97.

\bibitem[Yulin \em{et~al.}(2004)Yulin, Skryabin, and Russell]{Yulin2004}
Yulin, A.V.; Skryabin, D.V.; Russell, P.S.J.
\newblock {Four-wave mixing of linear waves and solitons in fibers with
  higher-order dispersion}.
\newblock {\em Opt. Lett.} {\bf 2004}, {\em 29},~2411--2413.

\bibitem[Efimov \em{et~al.}(2004)Efimov, Taylor, Omenetto, Yulin, Joly,
  Biancalana, Skryabin, Knight, and Russell]{Efimov2004}
Efimov, A.; Taylor, A.J.; Omenetto, F.G.; Yulin, A.V.; Joly, N.Y.; Biancalana,
  F.; Skryabin, D.V.; Knight, J.C.; Russell, P.S.
\newblock {Time-spectrally-resolved ultrafast nonlinear dynamics in small-core
  photonic crystal fibers: Experiment and modelling}.
\newblock {\em Opt. Express} {\bf 2004}, {\em 12},~6498--6507.

\bibitem[Skryabin and Yulin(2005)]{Skryabin2005}
Skryabin, D.V.; Yulin, A.V.
\newblock {Theory of generation of new frequencies by mixing of solitons and
  dispersive waves in optical fibers}.
\newblock {\em Phys. Rev. E} {\bf 2005}, {\em 72},~016619.

\bibitem[Zaitsev(2001)]{Zaitsev2001a}
Zaitsev, A.M.
\newblock {\em {Optical Properties of Diamond}}; Springer Berlin Heidelberg:
  Berlin/Heidelberg, Germany,  2001; p. 502.

\bibitem[Tyrrell \em{et~al.}(2005)Tyrrell, Kinsler, and New]{Tyrrell2005}
Tyrrell, J.C.A.; Kinsler, P.; New, G.H.C.
\newblock {Pseudospectral spatial-domain: A new method for nonlinear pulse
  propagation in the few-cycle regime with arbitrary dispersion}.
\newblock {\em J. Mod. Opt.} {\bf 2005}, {\em 52},~973--986.

\bibitem[Conforti \em{et~al.}(2013)Conforti, Marini, Tran, Faccio, and
  Biancalana]{Conforti2013}
Conforti, M.; Marini, A.; Tran, T.X.; Faccio, D.; Biancalana, F.
\newblock {Interaction between optical fields and their conjugates in nonlinear
  media}.
\newblock {\em Opt. Express} {\bf 2013}, {\em 21},~31239--31252.

\bibitem[Couairon \em{et~al.}(2011)Couairon, Brambilla, Corti, Majus, {de J.
  Ram{\'{i}}rez-G{\'{o}}ngora}, and Kolesik]{Couairon2011a}
Couairon, A.; Brambilla, E.; Corti, T.; Majus, D.; {de J.
  Ram{\'{i}}rez-G{\'{o}}ngora}, O.; Kolesik, M.
\newblock {Practitioner's guide to laser pulse propagation models and
  simulation}.
\newblock {\em  Eur. Phys. J. Spec. Top.} {\bf 2011}, {\em
  199},~5--76.

\bibitem[Boyd(2008)]{Boyd2008b}
Boyd, R.W.
\newblock {\em {Nonlinear Optics}}; Academic Press: New York, US,  2008.

\bibitem[Rubino \em{et~al.}(2012)Rubino, Lotti, Belgiorno, Cacciatori,
  Couairon, Leonhardt, and Faccio]{Rubino2012}
Rubino, E.; Lotti, A.; Belgiorno, F.; Cacciatori, S.L.; Couairon, A.;
  Leonhardt, U.; Faccio, D.
\newblock {Soliton-induced relativistic-scattering and amplification}.
\newblock {\em Sci. Rep.} {\bf 2012}, {\em 2},~932.

\bibitem[Conforti \em{et~al.}(2014)Conforti, Baronio, and Trillo]{Conforti2014}
Conforti, M.; Baronio, F.; Trillo, S.
\newblock {Resonant radiation shed by dispersive shock waves}.
\newblock {\em Phys. Rev. A} {\bf 2014}, {\em 89},~013807.

\end{thebibliography}
%\renewcommand\bibname{References}

%%%%%%%%%%%%%%%%%%%%%%%%%%%%%%%%%%%%%%%%%%
\end{document}